\documentclass{sigchi}

\toappear{}
\usepackage{balance}  % to better equalize the last page
\usepackage{graphics} % for EPS, load graphicx instead
\usepackage{times}    % comment if you want LaTeX's default font
\usepackage{url}      % llt: nicely formatted URLs
\usepackage{booktabs}
\usepackage[usenames,dvipsnames]{xcolor}
\graphicspath{{}}

\makeatletter
\def\url@leostyle{%
  \@ifundefined{selectfont}{\def\UrlFont{\sf}}{\def\UrlFont{\small\bf\ttfamily}}}
\makeatother
\def\pprw{8.5in}
\def\pprh{11in}

\usepackage[pdftex]{hyperref}
\hypersetup{
pdftitle={Exploring Cyberbullying and Other Toxic Behavior in Team Competition Online Games},
pdfauthor={Haewoon Kwak, Jeremy Blackburn, and Seungyeop Han},
pdfkeywords={League of Legends, online video games, team competition, toxic behavior, cyberbullying, trolling, crowdsourcing, MOBA},
bookmarksnumbered,
pdfstartview={FitH},
colorlinks,
citecolor=black,
filecolor=black,
linkcolor=black,
urlcolor=black,
breaklinks=true,
}

\newtheorem{hyp}{Hypothesis} 
\newtheorem{hypothesis}{Hypothesis}[hyp]

\def\Hnospace~{H{}}

\begin{document}

\title{Exploring Cyberbullying and Other Toxic Behavior in Team Competition Online Games \\
{\LARGE [Please cite the CHI'15 version of this paper]}}

\numberofauthors{3}

\author{
  \alignauthor Haewoon Kwak\\
    \affaddr{QCRI}\\
    \affaddr{Doha, Qatar}\\
    \email{hkwak@qf.org.qa}
  \alignauthor Jeremy Blackburn\\
    \affaddr{Telefonica Research}\\
    \affaddr{Barcelona, Spain}\\
    \email{jeremyb@tid.es}
  \alignauthor Seungyeop Han\\
    \affaddr{University of Washington}\\
    \affaddr{Seattle, WA, USA}\\
    \email{syhan@cs.washington.edu}
}

\maketitle

\begin{abstract}
In this work we explore cyberbullying and other toxic behavior in team competition online games.
Using a dataset of over 10 million player reports on 1.46 million toxic players along with corresponding crowdsourced decisions, we test several hypotheses drawn from theories explaining toxic behavior.
Besides providing large-scale, empirical based understanding of toxic behavior, our work can be used as a basis for building systems to detect, prevent, and counter-act toxic behavior.
\end{abstract}


% \category{K.8}{Personal Computing}{General}[Games]

% \setcounter{section}{-1}

\section{Executive Summary}

The increasing prevalence of computer mediated communication (CMC) has brought with it a host of undesirable behavior.
In particular, cyberbullying and other toxic (i.e., ``bad'') behavior has become exceedingly problematic.
One of the most popular forms of CMC are online video games, which enable real time interaction between hundreds of millions of players across the world.
The unique elements of \emph{competitive} online games makes players particularly vulnerable to the exhibition of, and negative effects from, cyberbullying and toxic behavior.

In this paper, we analyze a few million reports on toxic behavior from the most popular online game in the world, the League of Legends (LoL).
We find that:
\begin{enumerate}
  \item Players are surprisingly not engaged in actively reporting toxic behavior.
  \item Engagement can be significantly increased via explicit pleas to report.
  \item There are significantly varying perceptions of what constitutes toxic behavior between those that experienced it and neutral 3rd party ``judges.''
  \item There are biases with respect to reporting allies vs. enemies.
  \item There are significant cultural differences in perceptions concerning toxic behavior.
  \item The result of a match is significantly linked to the appearance of toxic behavior.

Our findings suggest several avenues for designers of video games in particular, and CMC systems in general.
\end{enumerate}

% See: \url{http://www.acm.org/about/class/1998/}
% for more information and the full list of ACM classifiers
% and descriptors. \newline
% \textcolor{red}{Optional section to be included in your final version, 
% but strongly encouraged. On the submission page only the classifiers’ 
% letter-number combination will need to be entered.}

\section{Introduction}

With the remarkable advances from isolated console games to massively multi-player online role-playing games (MMORPG), the online gaming world provides yet another place where people interact with each other. 
The main reasons that researchers pay attention to online games are 1)~that the purpose of actions is relatively clear, and 2)~that actions are quantifiable.
A wide range of predefined actions for supporting social interaction (e.g., friendship, communication, trade, enmity, aggression, and punishment) reflects either positive or negative connotations among game players~\cite{Szell10}, and they are unobtrusively recorded by game servers.
These rich electronic footprints enable and encourage research of social dynamics~\cite{ducheneaut2007life,Huang13,Szell10}.
 
There is no end to the growth of online gaming in sight.
For example, in 2011 the First Dota 2 International Tournament had a prize pool of \$1,600,000.
In 2014, the prize pool started again at \$1.6M, but, a \emph{crowdfunding} mechanism brought it up to just under \$11M.
For comparison, the winners of the 2014 International received \$5M (\$1M for each team member), while the winner of the 2014 US Open Men's tennis tournament received \$3M.
In other words, video games have reached a level of sophistication and competitiveness that not only can gamers make a living playing them, they can make quite a comfortable living. 

While the widely adopted game design element of competition increases enjoyment~\cite{Vorderer03}, it has also led to an increasing concern about negative behavior.
In online gaming, negative behavior, such as cyberbullying~\cite{Smith08}, griefing~\cite{Foo04}, mischief~\cite{Kirman12}, and cheating~\cite{Blackburn12} are often grouped together and called \emph{toxic behavior}.
Unfortunately, the definition of toxic behavior is often unclear~\cite{Chesney09} due to differences in expected behavior, customs, rules, and ethics across games~\cite{Warner05}.
Such ambiguity and subjective perception of griefing make griefers themselves sometimes fail to recognize what they did~\cite{Lin05}.

To make matters worse, since online games are very popular in the younger generation~\cite{demographics@ign}, instances of cyberbullying can cause far reaching problems.
In general, cyberbullying is associated with depression, anxiety, and has been shown to result in drastic actions such as suicide in several well publicized cases~\cite{suicideMegan}.
With the amount of time and energy players invest into games, victims of toxic behavior are likely to feel emotional effects that persist to the real-world.

In this paper, we make use of a large-scale dataset capturing millions of instances of toxic behavior perpetuated by hundreds of thousands of accused toxic players from the League of Legends (LoL)\footnote{http://www.leagueoflegends.com}, the world's most popular online game~\cite{Gaudiosi12}.
The richly detailed cases are augmented by crowdsourced decisions on whether or not the accused was in fact toxic.
Drawing from sociology and psychology literature, we explore toxic behavior through the lens of competitive online games.

\section{Background}

In this section we begin with a basic description of LoL and the Tribunal, its crowdsourcing platform for addressing toxic behavior.
We highlight the differences between LoL and other online games from the perspective of team formation, the goal of the game, and the common game modes.
We then move on to how Riot Games attempts to detect toxic players and which behavior is considered toxic.
Lastly, we explain how the Tribunal system works.  

\subsection{League of Legends as a Team Competition Game}

The League of Legends (LoL), a Multiplayer Online Battle Arena (MOBA), is arguably the most popular online game in the world.
The developer, Riot Games, recently announced there are 67 million players per month, 27 million players per day, and over 7.5 million concurrent players at peak times~\cite{27Mperday}.
LoL is an online team competition game whose goal is to penetrate and destroy the enemy's central base, called the Nexus.  
In contrast to massively multiplayer online role-playing game (e.g., World of Warcraft), LoL is a match-based team competition game, where a single match usually takes around 30 to 40 minutes.  
Although LoL provides a few different modes of matches, all the modes are competitions between two teams.  
In the most popular game mode, each team has five members.  
A player is given the option to form a team with friends before the match.
Otherwise, the player is randomly assigned to a team with strangers who have similar skill levels.

\subsection{Player Reports on Toxic Behavior}

\begin{figure} [hbt!]
  \begin{center}
  \includegraphics[width=80mm]{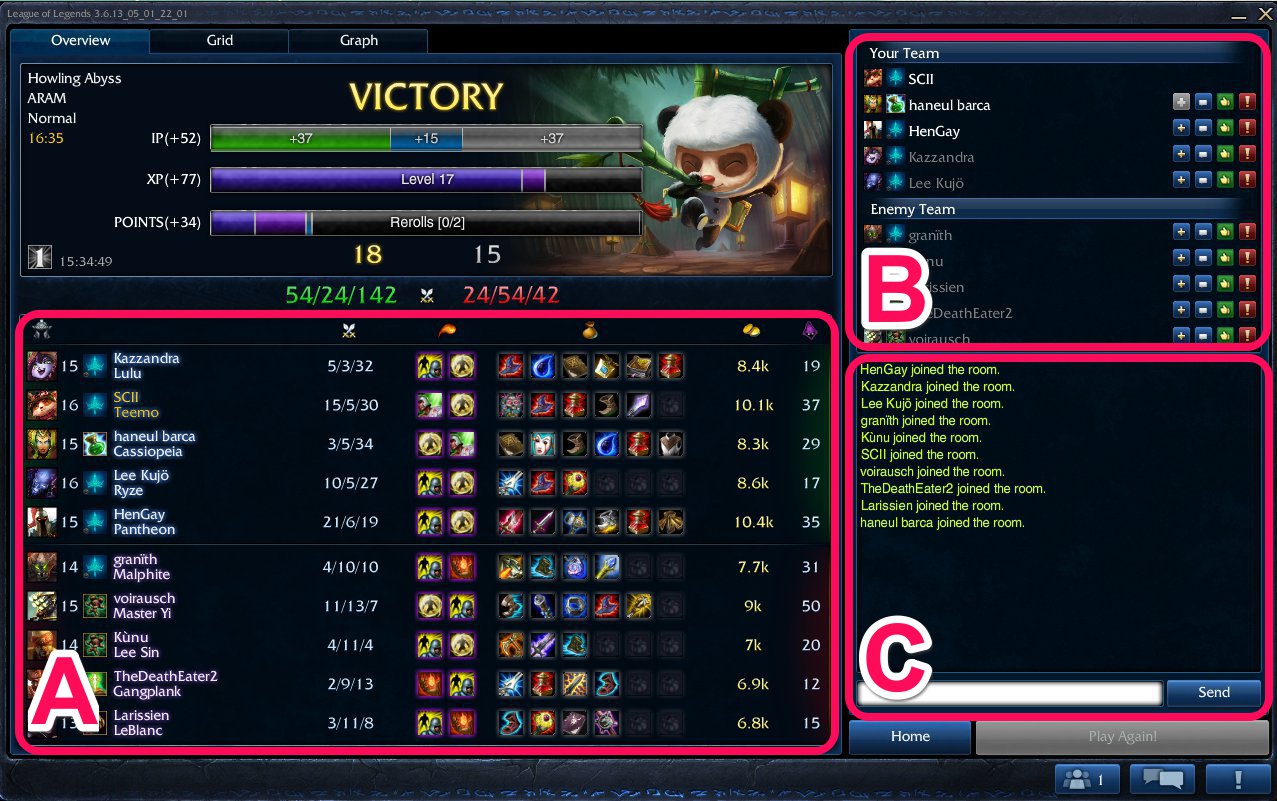}
  \caption{LoL scoreboard after a match.}  
  \label{fig:scoreboard}
  \end{center}
\end{figure}

After a match, players see a scoreboard as in \autoref{fig:scoreboard}.
In \autoref{fig:scoreboard}, (A) is the match summary, (B) lists players who played the game together, and (C) is a chat window. 
Users can report toxic players by clicking the rightmost red button in (B).  
Each player can submit one report for every other player per game. 
We stress that player reports can only be submitted \emph{after} the match is completed, and at a later date reviewers vote as described in the next section.
Therefore, player reports are not considered a strategic element during the match itself.

When reporting, players choose from several predefined categories of toxic playing: assisting enemy team, intentional feeding $[$suicide$]$, offensive language, verbal abuse, negative attitude, inappropriate name, spamming, unskilled player, refusing to communicate with team, and leaving the game$/$AFK $[$away from keyboard$]$.  Detailed explanations of each type of toxic playing are given in \cite{toxicplaying@support}.  Each report is submitted under one category. 
We further classify these categories as either cyberbullying (offensive language and verbal abuse) or domain specific toxicity (the remaining categories). 

While most of categories are self explanatory, two of the domain specific categories need additional explanation: intentional feeding and leaving the game.
Intentional feeding indicates that the player died to the opposing team on purpose, often repeatedly.
Leaving the game$/$AFK refers to instances when a user is inactive through the entire match. 
These two categories of toxic behavior usually make the opposing team stronger due to the game's design.

To understand the motivation behind intentional feeding and leaving the game, we present one common scenario where such toxic play happens. 
A LoL match is concluded when the Nexus of one team is destroyed.  
If one team is obviously at a disadvantage during the game, the team may give up the match by voting to ``surrender''.
However, surrender is accepted only when at least four out of five players on the same team agree to do so.
When the surrender vote fails, players who voted for surrender may lose interest in continuing the game.
Then, they might exhibit extreme behavior, such as intentional feeding or leaving the game$/$AFK, in an attempt to force the game to finish earlier or convince other players to cast a vote for surrender.

\subsection{LoL Tribunal as a Crowdsourcing System}

The LoL Tribunal is a crowdsourcing system to make decisions on whether reported players should be punished or not.
Only players that are reported more than a few hundred times are brought to the Tribunal~\cite{blackburn2014stfu}.
Once this threshold is reached, up to 5 randomly selected matches that an accused player was reported in are aggregated into a case.
Reviewers, who are LoL players with enough time invested to be considered ``experts,'' are then presented with a rich set of information about the case.
They see the full chat logs from each match, the performance (similar to the end score board) of each player in the game, the duration of the match, and the category (and optional comments) that the player was reported for.
To ensure an unbiased result, all players are anonymized in the match data.
We note that among 10 predefined categories of toxic behavior, the Tribunal excludes reports of unskilled player, refusing to communicate with team, and leaving the game.  

It is known that about 100-150 reviewers cast votes for a single case~\cite{blackburn2014stfu}, and a majority voting scheme is used to reach a verdict.
Guilty verdicts can result in a ban for 1 week, a few months, and sometimes longer suspensions/permanent bans.
E.g., professional players have received lengthy, career impacting suspensions due to Tribunal decisions~\cite{iwilldominate}.

\section{Data Collection}

We collect all information available in the Tribunal: player reports, the crowdsourced decision, and detailed match logs.  This rich collection is the basis for our quantitative analysis.  

Early studies on moral behavior in sports largely depend on self-reports~\cite{ommundsen2003perceived}, but they suffer in quality and scalability from bias in the human recall process~\cite{marsden1984measuring}.  
Furthermore, self-reports in particular can be influenced by social desirability, the tendency to behave in a way that is socially preferable~\cite{kavussanu2003participation}.
In this sense, reports submitted by 3rd parties, as in the Tribunal, are more reliable since they eliminate the bias of self-report.

Another widely used method is asking individuals about hypothetical situations, but it also has the same social desirability bias, and additionally has limitations due to the specificity of scenarios and scalability~\cite{paulhus1991measurement}.  
Our huge collection of witness reports on \emph{observed} toxic behavior helps address these problems.
Furthermore, crowdsourced decisions offer an even more objective viewpoint on reported toxic behavior.  

Riot Games divides the world into several regions and maintains dedicated servers for each region.
We focus on three regions, North America (NA), Western Europe (EUW), and South Korea (KR) by considering representativeness of cultural uniqueness and familiarity to the authors\footnote{Well over 4,500 combined hours in MOBAs.}.
Although a player may connect to servers operated in different regions, a further distance between a player and a server usually results in increased latency in Internet connections, sacrificing the quality of responsiveness and interactiveness of online games.  
We thus reasonably assume most players connect to the servers corresponding to their real-world region for the best quality of service.

In April 2013, we collected about 11 million player reports from 6 million matches on 1.5 million potentially toxic players across three regions.
We collect all available data from the servers and summarize it in \autoref{fig:data_collection}.
We first note that the KR portion of our dataset is smaller than other regions because the KR Tribunal started in November 2012 but the EUW and NA Tribunals started in May 2011.
Next, since player reports are internally managed, it is not easy to measure our dataset's completeness.
However, as reviewers can see their past votes and final decisions at a later time, it makes sense that no votes or reports are removed from servers and we have what amounts to a complete picture of the Tribunal.

\begin{table} [hbt!]
% \vspace{-2mm}
\centering
% \small
\begin{tabular}{crrrr}
\toprule
  & \multicolumn{1}{c}{\bf NA} & \multicolumn{1}{c}{\bf EUW} & \multicolumn{1}{c}{\bf KR} & \multicolumn{1}{c}{\bf TOTAL} \\
\midrule
{\bf Player} & 590,311 & 649,419 & 220,614 & 1,460,344 \\
{\bf Match} & 2,107,522 & 2,841,906 & 1,066,618 & 6,016,046 \\
{\bf Report} & 3,441,557 & 5,559,968 & 1,893,433 & 10,898,958 \\
\bottomrule
\end{tabular}
\caption{Summary of our dataset.}
\label{fig:data_collection}
% \vspace{3mm}
\end{table}

\section{Research Questions and Hypotheses}

In this section we formulate our research questions on how a toxic player behaves as well as how other players react to the toxic player.  

At its core, LoL is a team competition game.
Although competition against human opponents provides a high degree of player enjoyment~\cite{Weibel08}, it can also result in toxic behavior.
The potential for toxic behavior can be further exacerbated by the anonymous aspects of LoL, as people do not feel accountable for their toxic behavior when anonymous and can actually increase their aggressive behavior~\cite{christopherson2007positive}.
We present several compelling theories from sociology and psychology that describe the ``whys'' of toxic behavior.
Our large-scale dataset gives the opportunity to apply these theories ``in the wild.''

We begin with an explanation of the bystander effect and the vague nature of toxic playing.
Next, we discuss the concepts of in-group favoritism and out-group hostility with respect to competition.
We then describe how intra-group conflict emerges in a team competition environment like LoL and related theories which indicate a socio-political effect on how toxic behavior is both expressed and perceived.
Finally, we consider the effects of team-cohesion on performance, which can provide insights into what might trigger toxicity in online games in particular.

% for removing dangling
% \newpage

\subsection{Bystander Effect and Vague Nature of Toxic Playing}

Our first research question is focused on how actively people report toxic playing.  It is essential to understand the effectiveness and design implications of a system for counter-acting toxic playing, such as the Tribunal.

\textbf{RQ1a:} How active are players in reporting toxic behavior?

This research question is heavily related to the social psychology concept of the bystander effect~\cite{barlinska2013cyberbullying}, which describes the tendency for observers to avoid helping a victim, particularly when they are immersed in a group.
Considering LoL's anonymous setting with ephemeral teams, the bystander effect can influence how other players react to a toxic player. 
If the bystander effect is valid in this setting, then most players in a match will \emph{not} report the toxic player even though they directly witnessed the abuse. 
We explore how actively players report their observation of toxic behavior via our dataset.

Interestingly, the bystander effect is known to be mitigated with explicit pleas for aid~\cite{latane1970unresponsive}.  We can thus draw a testable hypothesis.

\setcounter{hyp}{1}
\begin{hypothesis}
\label{hyp:request-for-report}
If there is a request asking to report toxic players, the number of reports is increased.
\end{hypothesis}

\textbf{RQ1b:} How does the vague nature of toxic playing affect tribunal decisions?

Additionally, players have different perceptions on the severity of toxic behavior and thus sometimes fail to recognize its presence~\cite{Lin05}.
This vague nature of toxic playing might affect both reporting and reviewing.
We expect that different perceptions between players who report toxic behavior and Tribunal reviewers will result in a number of pardons.

\subsection{In-group Favoritism and Out-group Hostility}

The next question we explore is related to the team-based, competitive gameplay of LoL.

\textbf{RQ2:} What is the difference between reporting behavior of the toxic player's teammates and his opponents?

In-group favoritism is simply the tendency of people to favor in-group members (e.g., teammates) over similarly likable out-group members (e.g., opponents) while \emph{disliking} out-group members when compared to similarly dislikable in-group members~\cite{turner1975social}.
This is also related to homophily~\cite{mcpherson2001birds}, which is the tendency for similar individuals to form relationships.

Deindividuation theory explains how an individual loses the concept of both self and  responsibility when immersed in a crowd~\cite{Diener79}, which provides a possible mechanism behind in-group favoritism.
This foundation meshes with the unique characteristics of computer-mediated communication (CMC): anonymity and reduced self-awareness~\cite{Chen09,Thompson96}.

Reicher et al.~\cite{Reicher95} proposed the Social Identity Model of Deindividuation Effects (SIDE) as a way of explaining effects that classic deindividuation theory could not.
In short, they discovered that simply being part of an anonymous crowd was insufficient impetus for displaying anti-normative behavior.
Instead, in-group identity alone becomes relevant only in comparison to a relevant out-group.
In other words, both anonymity \emph{and} context drive deindivduation.

Even though the groupings in LoL are ephemeral and mostly anonymous, 
the two teams are in direct competition and only one team can win:
the thrill of victory or the agony of defeat is shared by all members of a team.
This configuration catalyzes group identification and provides clear cut divisions of in- and out-groups in a context that is \emph{driven} by out-group hostility; i.e., players are literally attempting to defeat the opposing team.
Thus, we expect to see empirical evidence of in-group favoritism in reporting of some toxic behavior that equally affects both teams.

\setcounter{hyp}{2}
\setcounter{hypothesis}{0}
\begin{hypothesis}
\label{hyp:in-group-favoritism}
For toxic behavior that affects both teams equally, in-group members (teammates) are less likely to submit reports when compared to out-group members (opponents).
\end{hypothesis}

\subsection{Intra-group Conflicts and Socio-political Factors}

Although we expect to find evidence of in-group favoritism and out-group hostility, the team competition setting of LoL might also lead to substantial intra-group conflict.
Intuitively, it is easy to blame overall team performance on a single poorly performing player.
Unlike the aforementioned concept of in-group favoritism, the poorly performing player does not affect both teams equally because he makes the opponent team relatively stronger.  Thus, he becomes a target to blame.

According to the classification scheme of human society introduced by T{\"o}nnies~\cite{Tonnies12}, teams in LoL are closer to task-oriented associations (Gesellschaft) than the social community associations (Gemeinschaft).  
In task-oriented associations, the relationship among players is somewhat impersonal and social bonding does not necessarily exist.  
Thus, toxic players might not feel a sense of a team and feel no qualms about harassing teammates who are hurdles to winning rather than recognizing enemies for beating his team.

While it is difficult to make direct conclusions about intra-group conflict, LoL players are generally segregated by what amounts to socio-political regions.
The large world-wide user-base of LoL makes it feasible to study regional differences of such intra-group conflicts.  
Thus, leading to our next research question.

\textbf{RQ3:} What is the impact of socio-political factors on toxic behavior?

There are several studies to support that the degree to which bullying occurs is influenced by socio-political factors~\cite{aggression2011cyberbullying,Chee06}.
Chee reports on a unique Korean gaming culture called \emph{Wang-tta}~\cite{Chee06}.
In the context of gaming, \emph{Wang-tta} describes the phenomenon of ``isolating and bullying the worst game player in one's peer group.''
\emph{Wang-tta} is thought to be modeled after the similar Japanese term \emph{Ijime}~\cite{aggression2011cyberbullying} which describes the comfort members of collectivist societies feel from similarity and the abuse thrust upon those that are different.

Conversely, other socio-political regions such as North America and Western Europe tend to be individualistic, with a focus on relying on ones' self~\cite{hofstede2001culture}.
In such socio-political regions, while bullying of poorly performing players certainly occurs, there is less ingrained hostility towards another player's poor individual performance~\cite{Naito05}.
For example, consider the counterpart to the clich\'{e} ``there's no `I' in team,'' ``well there ain't no `WE' either.''
There is simply more focus on \emph{my} performance as opposed to \emph{our} performance.
If there are in fact socio-political factors at play, then we would expect to see this reflected in reports on toxic behavior from different regions, leading to a testable hypothesis.

\setcounter{hyp}{3}
\setcounter{hypothesis}{0}
\begin{hypothesis}
\label{hyp:punishment-cyberbullying-kr}
Due to a more group-success oriented socio-political environment, cyberbullying offenses are less likely to be punished in Korea than in other regions. 
\end{hypothesis}

Collectivist societies such as Korea and Japan have a desire towards similarity, while differences are a source of disparagement~\cite{Naito05}.
In addition, collectivism, by definition, places more emphasis on cooperation, the group goal, and a sense of belonging than individualism does~\cite{Cox91,Triandis88}.
That is, deliberately harming the group is anathema to the highly held social value of cooperation and is met with intense derision.
Thus, we propose the next two hypotheses from the perspective of reporters and reviewers.

\begin{hypothesis}
\label{hyp:reports-kr}
Reports on toxic behavior that largely affects the result of the match are more often submitted in Korea than in other regions.
\end{hypothesis}

\begin{hypothesis}
\label{hyp:punishment-feeding-kr}
Reports on behavior that largely affects the result of the match are more likely to be punished in Korea than in other regions.
\end{hypothesis}

\subsection{Team-cohesion and Performance}

The cohesion-performance relationship in sports has been studied for decades.
Several studies have confirmed a moderate but significant effect of cohesion on performance, with the effect varying according to the type of sport, the mechanism of team building, and the gender of players~\cite{Carron02,Mullen94}.
More generally, Felps et al. discuss how a single negative member can bring group-level dysfunction~\cite{felps2006how}.
The negative member violates interpersonal social norms and thus can lead to negative emotions and reduced trust among teammates.
These psychological states then trigger defensive reactions and influence the overall functioning of the group.
Intuitively, toxic behavior is likely to have a negative effect on team-cohesion, and thus performance, leading to our next research question.

\textbf{RQ4:} What is the relationship between toxic behavior, player reports, and team performance?

Weiner proposes attribution theory which states that individuals are likely to search for causal factors of failure, considering even innocuous factors as significant~\cite{weiner1980cognitive}.
Naquin and Tynan present a similar concept, the team halo effect~\cite{naquin2003team}, which describes the tendency of people to give credit for success to the team as a collective, but to blame other people for poor team performance.

The underlying cognitive process is known as counterfactual thinking~\cite{roese1997counterfactual}, which is when we create a mental simulation built on events that are contrary to the facts of what really happened.
If an individual deduces a high likelihood of changing the outcome to a more positive one via the mental simulation, then the difference between simulation and reality is identified as the causal factor for the negative outcome.

Generally speaking, counterfactual thinking helps us accurately identify causal factors of outcomes, but it can be biased by numerous factors~\cite{einhorn1986judging}, such as the halo effect.
For example, if one player makes a poor decision, then a toxic player might imagine what would have happened if that decision was not made, attributing the current state of the match to that singular incident regardless of any mistakes he or other players might have made.
Similarly, after the match is completed, a negative outcome and counterfactual thinking might lead bullied players to seek revenge on toxic players as a defensive reaction to perceived victimization and a method by which to gain some emotional satisfaction~\cite{denant2007punishment}.

Since theory indicates negative outcomes trigger both toxic behavior and attempts to punish said toxic behavior, but reviewers of Tribunal cases are unbiased, we propose the following hypotheses.

\setcounter{hyp}{4}
\setcounter{hypothesis}{0}
\begin{hypothesis}
\label{hyp:more-reports-from-defeats}
More reports come from matches where the accused was on the losing team.
\end{hypothesis}

Also, such aggressive reporting might be likely to be pardoned.

\begin{hypothesis}
\label{hyp:more-pardons-from-defeats}
There are more cases pardoned when the accused was on the losing team than on the winning team.
\end{hypothesis}

\section{Results}

\subsection{Bystander Effect and Vague Nature of Toxic Playing}

First, we look into how actively people report toxic playing.
When toxic play occurs in LoL, either 4 players (allies of the toxic player) or 9 players (all players but the toxic player) are exposed.
For instance, if a toxic player is verbally abusive or uses offensive language in the ally chat mode, it is only visible to his allies.
However, he can also use the all chat mode, exposing all 9 players to toxicity.
For the other categories, all players are exposed in essentially the same manner.

\begin{figure}[hbt!]
  \begin{center}
  \includegraphics[width=\columnwidth]{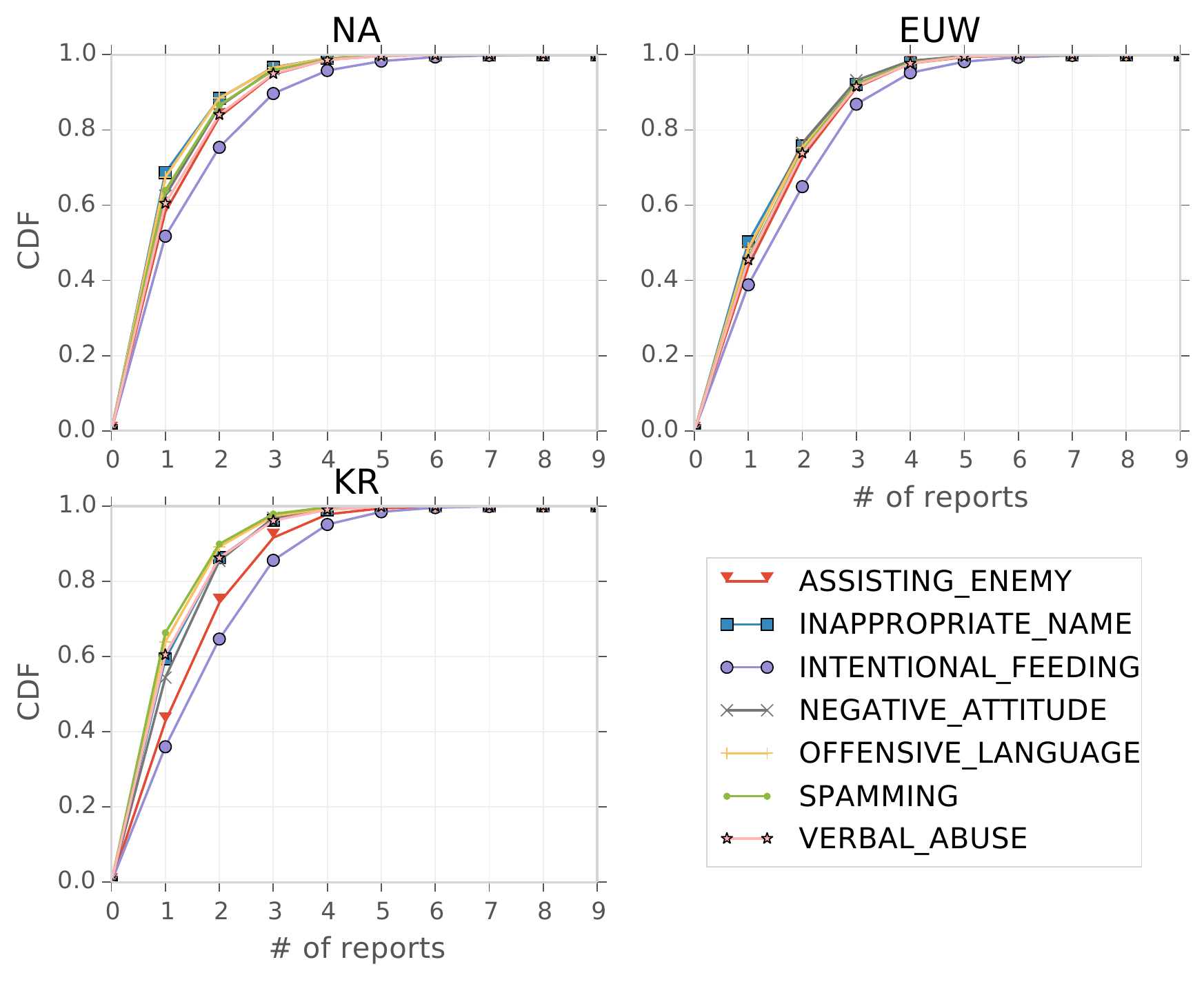}
  \caption{Distribution of the number of reports per match.}  
  \label{fig:number_of_reports_per_match}
  \end{center}
\end{figure}

Figure~\ref{fig:number_of_reports_per_match} plots the distribution of reports per match depending on the type of behavior reported for each region.
Across all regions, the mean and median number of toxic reports per match are 1.812 and 1, respectively\footnote{Every match in our dataset has at least one player report, as only players that have been reported appear in the Tribunal.}.
I.e., no more than 2 players per match report toxic playing on average. 
Overall, intentional feeding matches have the highest number of reports---about 50\% of cases have more than 1 report, with over 60\% for KR and EUW in particular---but the majority of matches have less than 3 reports.
This is extremely low compared to the number of exposed players.
I.e., on average, LoL players do not actively report toxic players.

To test \textit{[\autoref{hyp:request-for-report}] If there is a request asking to report toxic players, the number of reports is increased}, we look for situations where a request to report is sent to players on the enemy team; i.e., those that are not negatively affected by the toxic player, yet are likely to recognize his behavior and react if a plea is made.
An explicit request sent to the opposite team when intentionally feeding or assisting the enemy satisfies this condition.

To test this, we define two dichotomous variables: 1)~whether an explicit request exists, and 2)~whether members of the opposite team report intentional feeding or assisting the enemy.
We set the former variable to 1 when the word ``report'' appears in all chat, and 0 otherwise.
We then create a 2$\times$2 matrix of these variables.  
A Chi-Square test with Yates' continuity correction reveals that the percentage of reporting by enemies significantly differs with the existence of explicit requests ($\chi^2$(1, N = 580,480) = 194552.9, p $<$ .0001).

Surprisingly, through the odds ratio we discover that the probability of opponents reporting is 16.37 times higher when allies request a report.
This supports \autoref{hyp:request-for-report}: explicit requests to report toxic players highly encourage enemies to report toxic players \emph{even if toxic behavior is beneficial to the enemies}.
In other words, we find the bystander effect is neutralized via explicit requests for help. This is interesting because opposing players can benefit from toxic playing, and their typical behavior of not reporting is changed due to explicit requests.
This finding suggests that interaction design should actively encourage players to report others' toxic playing.

We next explore the possible impact of the vague nature of toxic playing on reporting and reviewing.
First, we look into an association between the recognizability of toxic playing and the number of reports.
Among the 7 types of toxic playing in LoL, intentional feeding and assisting the enemy team are, generally speaking, much more concrete expressions of toxicity than other types.
The average number of reports per match for these two categories is higher than that for other types, as shown in Table~\ref{tbl:report_per_match}. 
Both categories are consistently the first and the second ranked in terms of the average number of reports per match across all regions.
A Kruskal Wallis test revealed a significant effect for category of toxic behavior on the number of reports per match ($\chi^2$(6) = 117,399.1, p $<$ .0001).
Further post-hoc statistical testing using the pairwise Wilcoxon test with Bonferroni correction showed a significant difference between categories (p $<$ .0001).

% \vspace{-1mm}
\begin{table}[h]
\small
    \begin{center}
    \begin{tabular}{ c c c c c c c }
            %\hline
            \toprule
              A.E & I.N & I.F & N.A & O.L & S & V.A \\
            \midrule
            % Count & 45,762 &  133,948 & 46,186  & 78,793 \\
             1.876 & 1.602 & 2.091 & 1.696 & 1.691 & 1.769 & 1.740 \\
            \bottomrule
    \end{tabular}
    \end{center}
\caption{Avg. number of reports per match for each type of toxic playing.}
\label{tbl:report_per_match}
\end{table}
\vspace{-1mm}

\begin{figure} [hbt!]
  \begin{center}
  \includegraphics[width=\columnwidth]{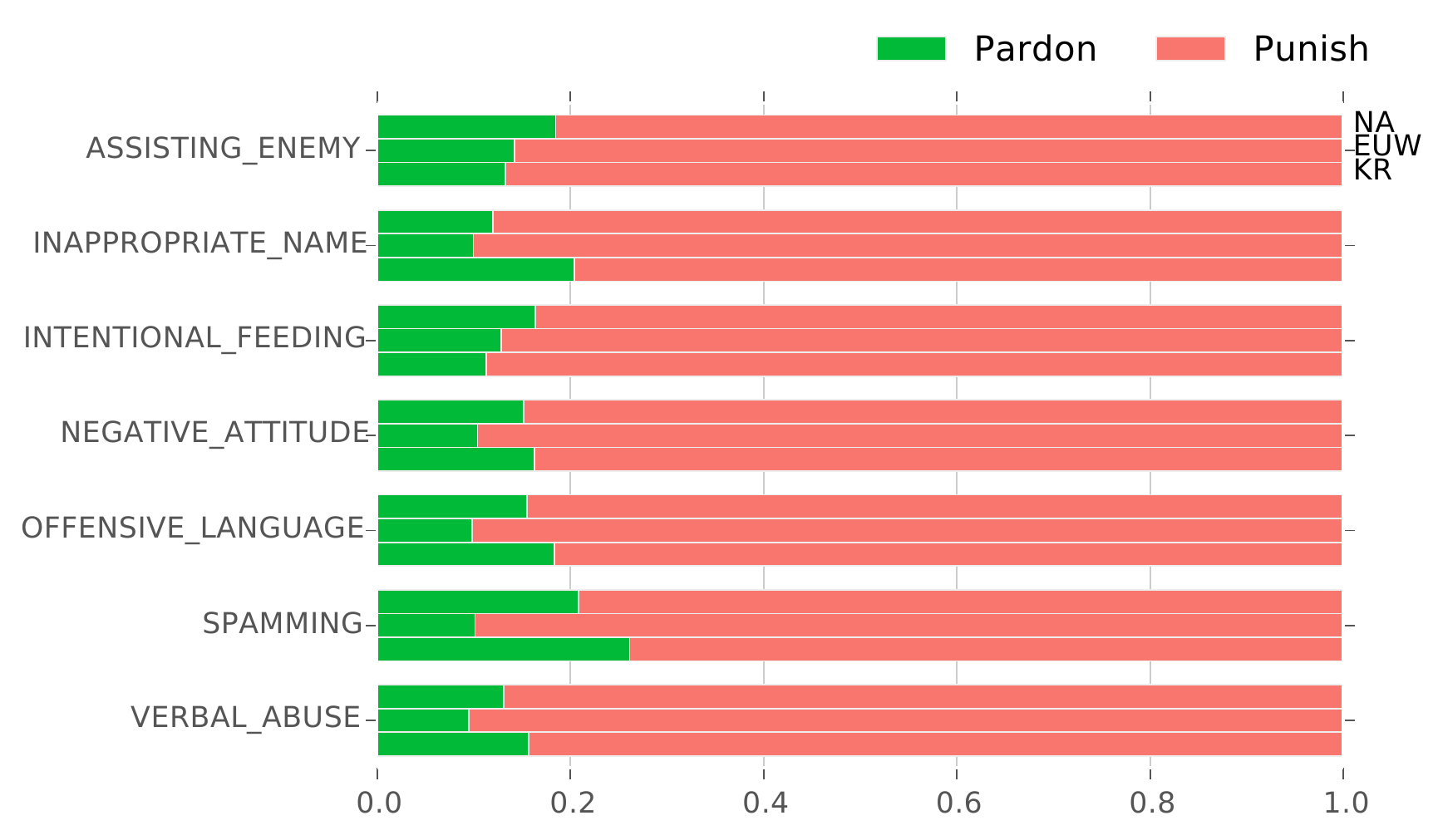}
  \caption{Proportion of decisions according to categories of toxic playing.}  
  \label{fig:proportion_of_decision}
  \end{center}
\end{figure}

Second, we look at how many reported toxic players are pardoned in Figure~\ref{fig:proportion_of_decision}.
Since players come to the Tribunal only after a few hundred reports against them, we do not expect to see a 50/50 punish/pardon ratio.
We obtained records for 477,383 punishments (80.9\%) and 112,930 pardons (19.1\%) for NA, 559,449 (86.0\%) and 90,966 (14.0\%) for EUW, and 187,253 (85.0\%) and 32,929 (15.0\%) for KR across all categories of toxic playing.
The highest pardoned ratio we found for a specific category\footnote{When multiple reports per accused player per game are submitted, we use the most common report type.}, spamming, is 26.1\% in KR.  
Being reported means that \emph{players} regard it as toxic playing, and a pardon means that \emph{reviewers} do not regard it as toxic.
Therefore, a 26\% pardoned rate shows that different perceptions exist: for 1 in 4 cases, reviewers do not find the player toxic.  
This high pardoned ratio confirms the impact of the vague nature of toxic playing in reviewing as well as in reporting.

\subsection{In-group Favoritism and Out-group Hostility}

Our next research question is understanding the difference between reporting behavior of the toxic player's teammates and his opponents. 
To test \textit{[\autoref{hyp:in-group-favoritism}] For toxic behavior that affects both teams equally, in-group members (teammates) are less likely to submit reports when compared to out-group members (opponents)}, we carefully revisit the definition of in-group favoritism.
The key concept of in-group favoritism is ``similarly'' likable or unlikeable members from the same group are more favorable.
We find this kind of relatively ``neutral'' toxic behavior from reports of inappropriate name.
The inappropriate name of a toxic player is visible to all players equally, and thus the impact of the inappropriate name is neutral to both teams.

\begin{figure} [hbt!]
  \begin{center}
  \includegraphics[width=\columnwidth]{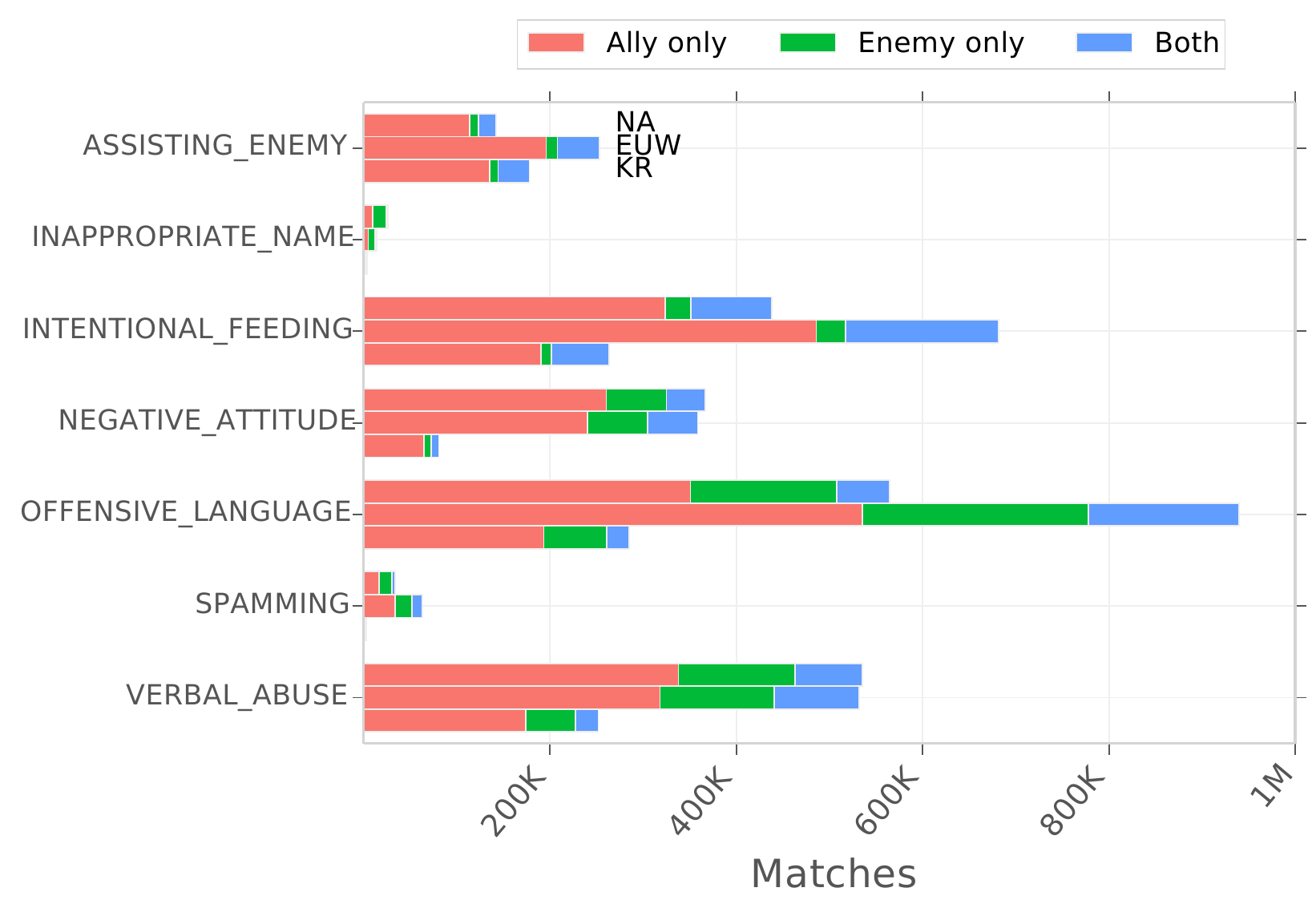}
  \caption{Number of matches that report each category of toxic behavior.}  
  \label{fig:ally_or_enemy}
  \end{center}
\end{figure}

In \autoref{fig:ally_or_enemy} we plot the number of matches reported for each category of toxic behavior per region.
As we hypothesized, the only category in which the number of reports from enemies is higher than from allies is inappropriate name (ally only: 16,339, enemy only: 23,966, across regions).
This indicates that allies are more forgiving, and thus less likely to report, than opponents in cases where the toxic player's offense is neutral to both teams.
This is exactly what in-group favoritism describes, and thus \autoref{hyp:in-group-favoritism} is supported.

\subsection{Intra-group Conflicts and Socio-political Factors}

We now explore our next research question on the relationship between intra-group conflicts and socio-political factors.
Figure~\ref{fig:ally_or_enemy} shows the number of matches that are reported due to each category of toxic behavior.
The horizontal bar for each category divided into three parts shows the number of matches reported by ally-only, enemy-only, and both, respectively from left to right. 
For instance, about 200 thousand matches are reported by ally-only due to assisting enemy, 12,106 matches by enemy-only, and 44,927 matches by both in EUW.  
We find that most reports come from allies rather from enemies as evidences of intra-group conflicts.

In Figure~\ref{fig:ally_or_enemy}, offensive language is the most reported toxic behavior in LoL and is highly reported by allies.  
The chat feature is designed for exchanging strategies and sharing emotions, and thus serves to foster a sense of belonging in the team.
In practice however, it becomes a channel for toxic players to harass other ally players~\cite{kwak2014linguistic}.

We additionally note that Riot Games disabled the all chat by default in newly installed clients since April 2012, while the Tribunal began in May 2011.
Although players can easily turn the all chat option on in the game configuration with a single click, we expect that it decreases the verbal interaction across teams after April 2012.

To confirm that this is not the reason that players are likely to report allies rather than opponents for verbal abuse or offensive language, we look into the temporal trend of the toxic reports for those two categories.
We divide toxic reports by a fixed-size time window, defined as 1,000 consecutive numeric identifiers of Tribunal cases.  
We compute the proportion of toxic reports that come from allies vs. those from opponents in each time window and find that, although it varies over time, the number of reports by allies is consistently higher than those by opponents.
This shows that frequent intra-group conflicts through chat are not artificial effects of the user interface.

\vspace{-1mm}
\begin{figure} [hbt!]
  \begin{center}
  \includegraphics[width=\columnwidth]{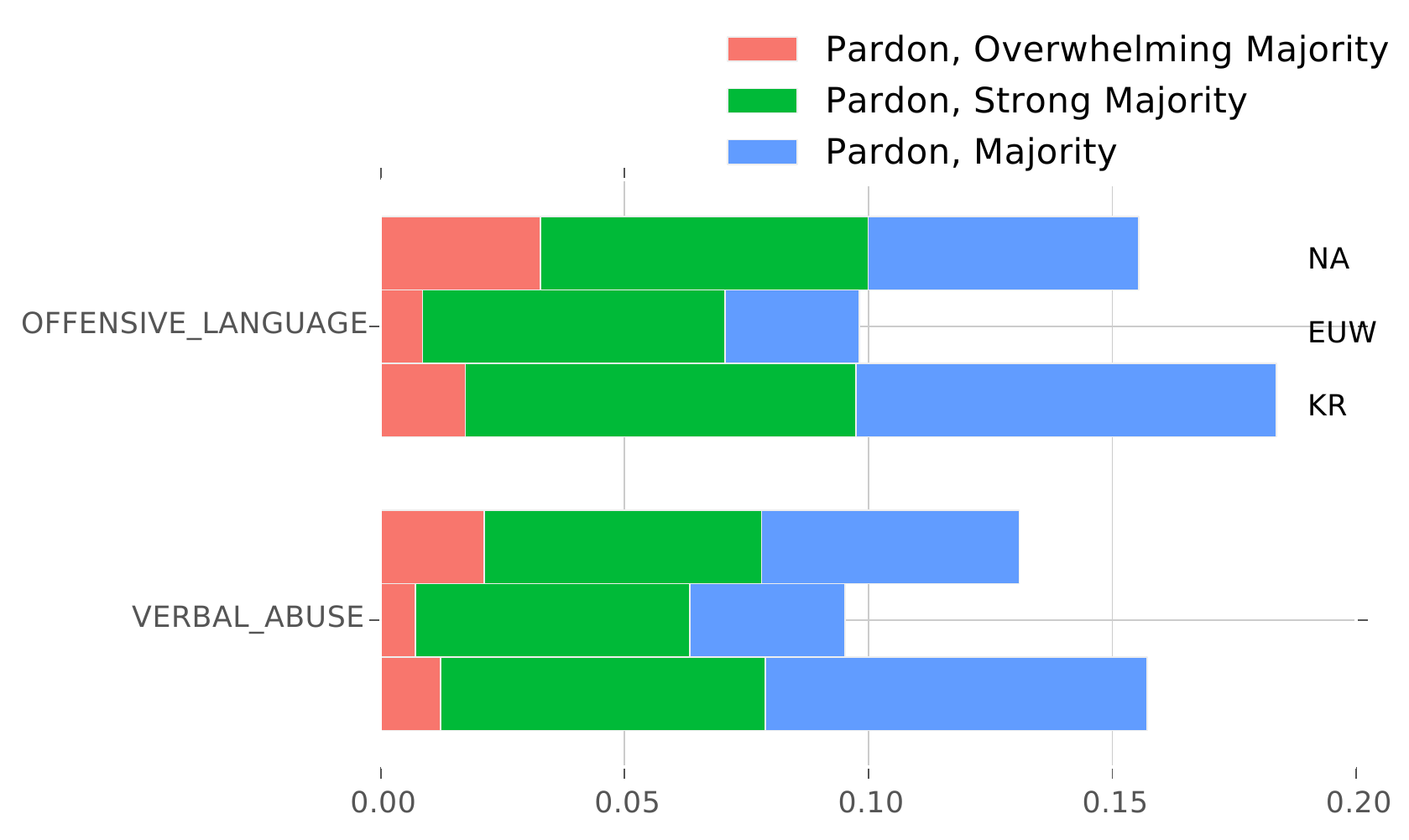}
  \caption{The percentage of cyberbullying reports that result in pardons for each region.}
  \label{fig:punishment-cyberbullying-kr}
  \end{center}
\end{figure}

To test \textit{[\autoref{hyp:punishment-cyberbullying-kr}] Due to a more group-success oriented socio-political environment, cyberbullying offenses are less likely to be punished in Korea than in other regions}, we examine the rate of pardons for cyberbullying offenses in the different regions.

\autoref{fig:punishment-cyberbullying-kr} plots the percentage of cyberbullying reports (offensive language and verbal abuse) that result in pardons for each region.
17.1\% of such reports are pardoned in KR, compared to 14.3\% in NA and 9.7\% in EUW. A Chi-Square test with Yates’ continuity correction reveals the effect of region on pardons in those categories is significant ($\chi^2$(2, N = 3,108,172) = 24123.3, p $<$ .0001).  Thus \autoref{hyp:punishment-cyberbullying-kr} is supported.

As we previously noted, a likely explanation for this is due to the \emph{Wang-tta} concept in KR.
Particularly invasive in gaming culture, \emph{Wang-tta} probably leads to reviewers empathizing not with the cyberbullying victim, but rather the alleged toxic player who verbalized his displeasure with the victim's performance.

To test \textit{[\autoref{hyp:reports-kr}] Reports on toxic behavior that largely affects the result of the match are more often submitted in Korea than in other regions}, we compare the percentage of reports for intentional feeding or assisting enemy across regions since such toxic behavior directly contradicts the group-success goal.   

We find support for \autoref{hyp:reports-kr} by looking at the mean and median of reports for assisting enemy or intentional feeding coming from teammates (1.482 and 1, 1.714 and 2, and 1.75 and 2 for NA, EUW, and KR, respectively).
A Kruskal-Wallis test reveals the effect of region on reports coming from teammates is significant ($\chi^2$(2) = 31175.43, p $<$ .0001).
A post-hoc test using Mann-Whitney tests with Bonferroni correction confirms the significant differences between NA and EUW (p $<$ .0001, r = .129), between NA and KR (p $<$ .0001, r = .148), and between EUW and KR (p $<$ .0001, r = .02).

\begin{figure} [hbt!]
\vspace{-1mm}
  \begin{center}
  \includegraphics[width=\columnwidth]{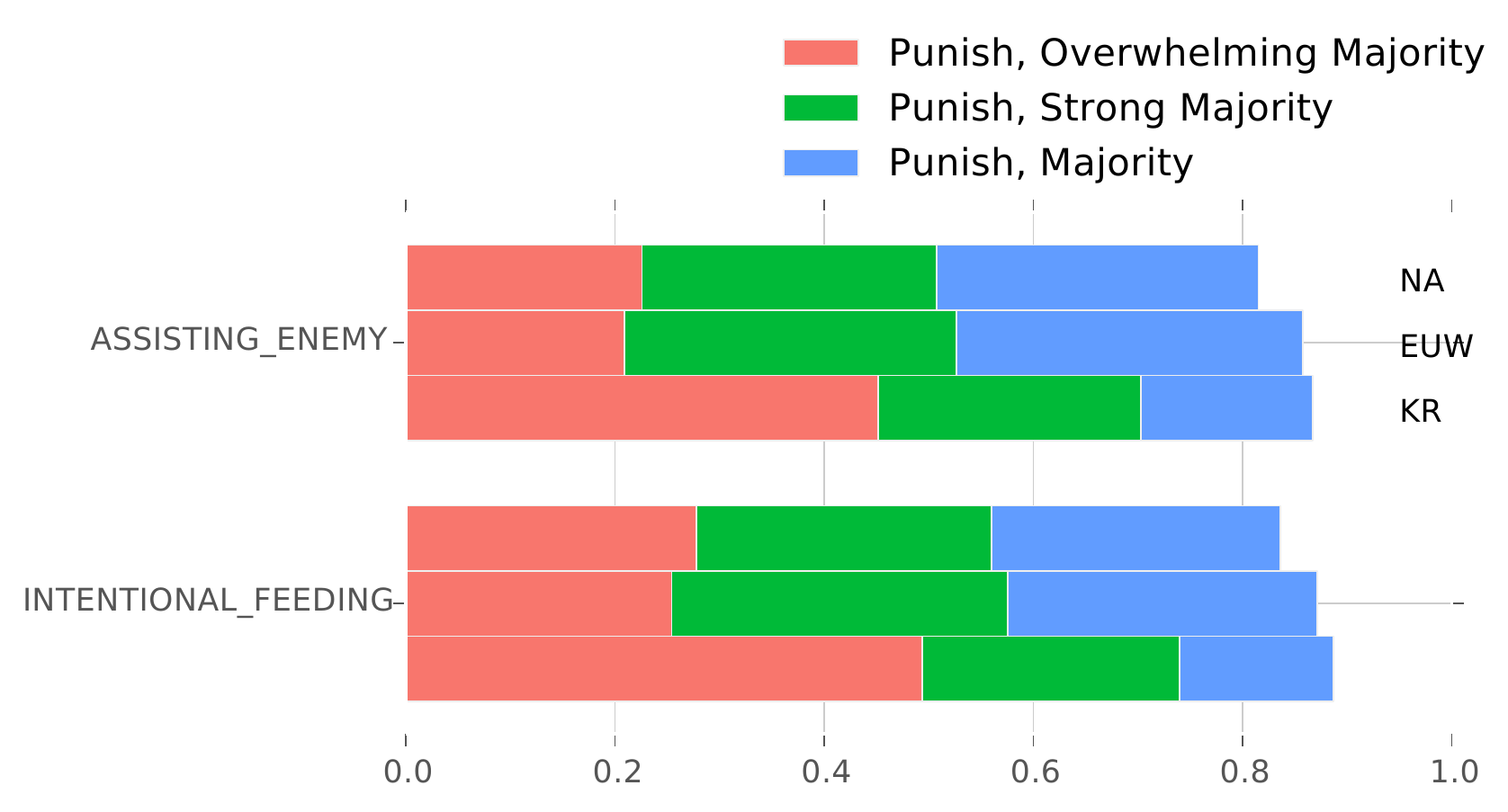}
  \caption{The percentage of reports for behavior that directly affects the result of the match for each region that results in a punishment.}
  \label{fig:punishment-feeding-kr}
  \end{center}
\vspace{-2mm}
\end{figure}

We move on to testing \textit{[\autoref{hyp:punishment-feeding-kr}] Reports on behavior that largely affects the result of the match are more likely to be punished in Korea than in other regions}.  
We plot the percentage of punish decisions for behavior that directly affects the result of the match for each region in \autoref{fig:punishment-feeding-kr}.
While such offenses are heavily punished in each region (over 80\%), we do find a difference.
Korean reviewers are more likely to perceive assisting enemy and intentional feeding as severe toxic playing with much higher levels of agreement than other regions (for Punish, Overwhelming Majority, KR: 48\%, NA: 27\%, EUW: 24\%), which is affirmative support for \autoref{hyp:punishment-feeding-kr}.
A Chi-Square test with Yates’ continuity correction reveals the effect of region on the percentage of Punish, Overwhleming Majority in those categories is significant ($\chi^2$(2, N = 1,955,297) = 83593.08, p $<$ .0001). %\hw{add some numbers to test hypothesis.}

\subsection{Team-cohesion and Performance}

\begin{figure}[hbt!]
  \begin{center}
  \includegraphics[width=\columnwidth]{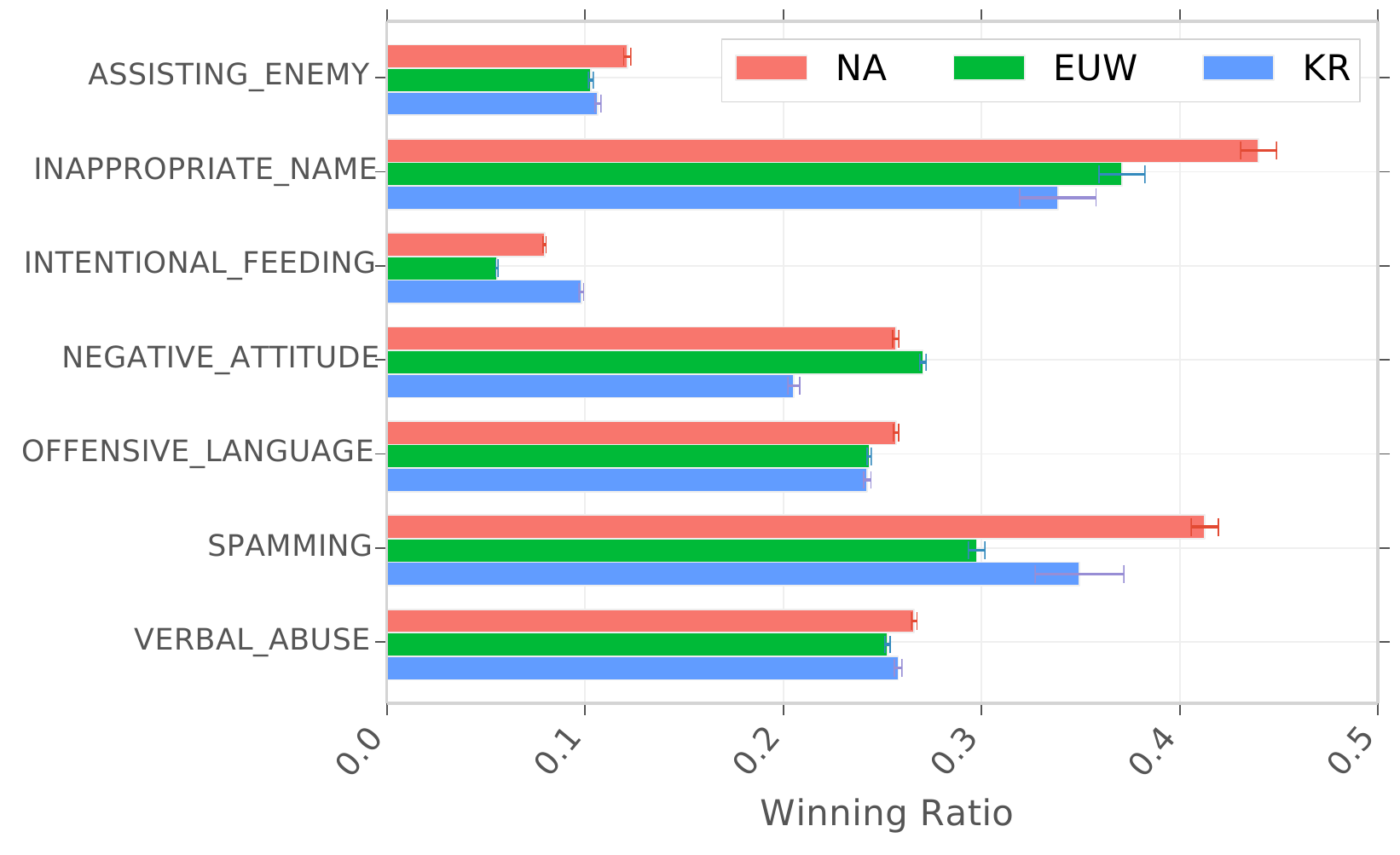}
  \caption{The winning ratio for each category of toxic behavior with 95\% confidence interval.}  
  \label{fig:winning_ratio}
  \end{center}
\vspace{-2mm}
\end{figure}

To test \textit{[\autoref{hyp:more-reports-from-defeats}] More reports come from matches where the accused was on the losing team}, 
we plot the winning ratio with 95\% confidence interval for the different categories of toxic play reported in \autoref{fig:winning_ratio}.
From the figure, it is immediately apparent that winning ratio is clearly below than 50\%, even though LoL uses a match making system similar to Elo ranking~\cite{elo1978rating} that attempts to match players in a manner where there is a 50\% chance of winning for each team.
In particular, we see that the winning ratio for intentional feeding and assisting the enemy are extremely low (under 15\% for both).
Therefore, more reports come from losing teams so that the observed winning ratio is less than 0.5 and \autoref{hyp:more-reports-from-defeats} is supported.

As mentioned in the research questions section, lower-team cohesion leads to lower performance.
Alternatively, a poor performance might trigger toxic behavior like cyberbullying and also might be a cause for the high number of reports involving a losing team.
Cyberbullying offenses are explained by attribution theory in that when a toxic player recognizes a poor performance (e.g., even though the match is not decided, his team is losing) he looks for someone other than himself to place blame.
Related to this, for example in the spamming category, players that were on the losing side of the match might attempt to attribute the loss to a another member of the team and attempt to punish him via the reporting system.

\vspace{-2mm}
\begin{figure} [hbt!]
  \begin{center}
  \includegraphics[width=70mm]{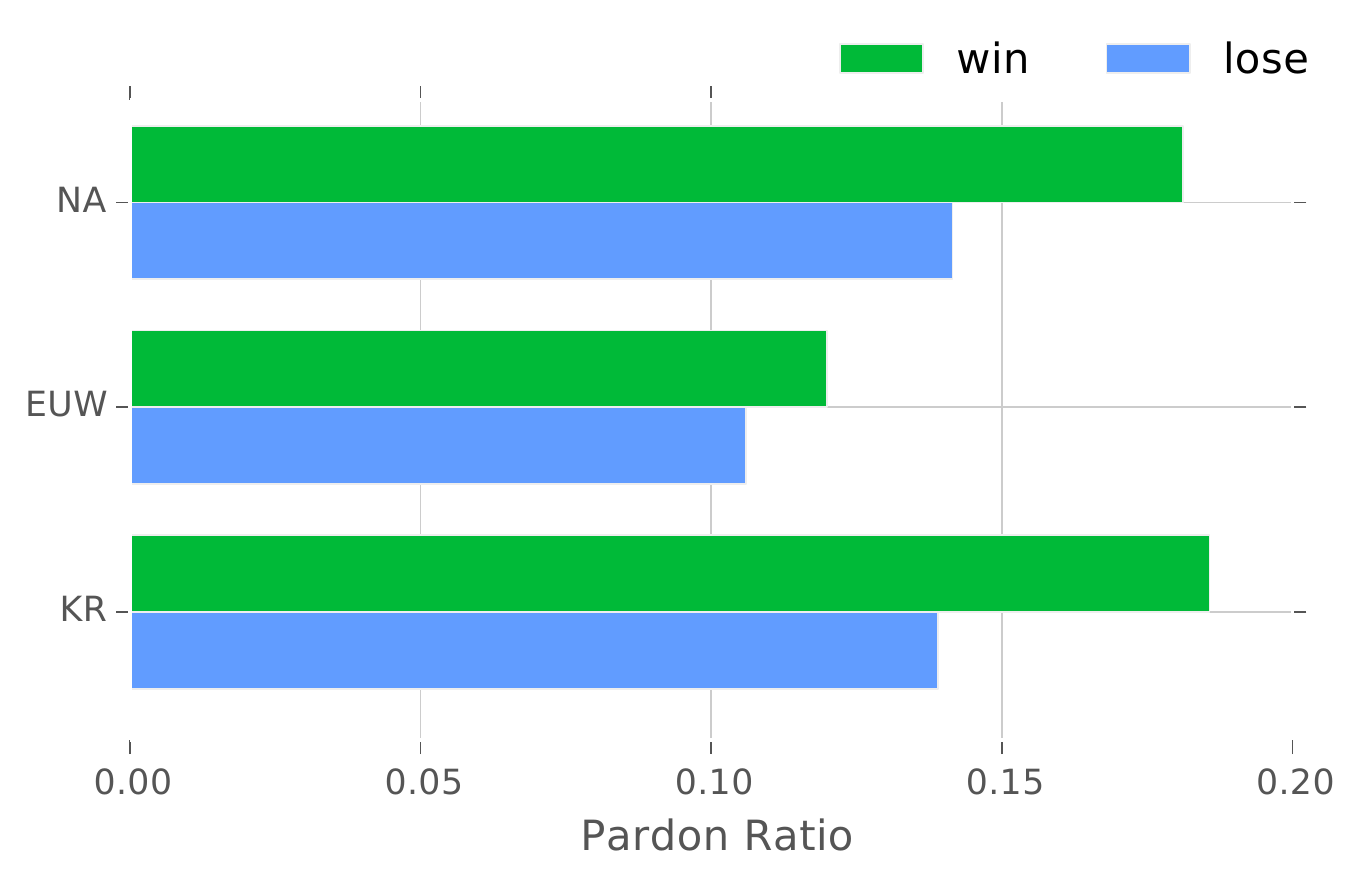}
  \caption{Pardoned ratio when losing vs. winning.}
  \label{fig:more-pardons-from-defeat}
  \end{center}
\end{figure}

To test \textit{[\autoref{hyp:more-pardons-from-defeats}] There are more cases pardoned when the accused was on the losing team than on the winning team}, we plot the pardoned ratio when the accused toxic player was on the winning or losing side.  
\autoref{fig:more-pardons-from-defeat} plots the breakdown of pardon and punish decisions as a function of whether the accused toxic player was on the winning or losing side.
If people regard somewhat innocuous factors as significant reasons of defeat and report it as toxic behavior, the pardoned ratio for defeats will be higher than that for wins.
As seen in \autoref{fig:more-pardons-from-defeat} however, we find that the proportion of being pardoned by crowds when losing is less than that when winning. 
Thus, \autoref{hyp:more-pardons-from-defeats} is \emph{not} supported, and in fact the opposite is the case.

\section{Discussion and Conclusion}

In this work we explored toxic playing and the reaction to it via crowdsourced decisions using a few million observed reports.
Our large-scale dataset enables an opportunity to explore several compelling theories from sociology and psychology that discuss the ``whys'' of toxic behavior.

We first showed that there is relatively low participation in reporting toxic behavior and reconfirmed the impact of anonymity in CMC and cyberspace.
This finding underlines the difficulties in relying solely on voluntary player reports.
For example, on Facebook there is a button to report a comment that violates cyber-etiquette.
The low degree of participation for social control we found raises a fundamental question about the design considerations of such report-based systems: if relatively few ``victims'' voluntarily report such behavior, how effective can it truly be?
Interestingly, our finding that an explicit request to report toxic behavior significantly increases the likelihood of reporting indicates that actively encouraging reporting should be considered in the design of systems to address toxic behavior.

Next, we examined the vague nature of toxic behavior.
This issue is repeatedly observed in online games~\cite{Lin05}.
With over 10\% of cases being pardoned in the Tribunal, it is clear that crowdsourcing using experienced players is useful in protecting innocent victims who are wrongly reported by other players due to poor gaming skills or aggressive (but not toxic) linguistic behavior.
The Tribunal could further improve the quality and efficiency of crowdsourced workers via several proposed mechanisms for quality control in crowdsourced systems and augmentation with machine-learning solutions~\cite{blackburn2014stfu}.

We then moved on to how group setting might influence reporting behavior.
We quantitatively show that in-group favoritism and out-group hostility increase or decrease the willingness to report.
This is rooted in competition among players.
Since competition is a common game design element for enjoyment~\cite{Vorderer03}, our findings are applicable to most online games.
Even though a given game might not have a form of team competition, sense of belonging is frequently derived from a small group of players, called guilds or parties in MMORPGs~\cite{ducheneaut2007life}.
In a broad sense, homophily can appear not only in an explicit group but also in an implicit group~\cite{mcpherson2001birds}.
The current work portends possible bias of observed reports in such settings.  

Most online games allow interactions between players.
In this environment toxic playing is a serious issue that degrades user experience. Our work offers understanding of toxic playing and its victims based on LoL data, but much of the mechanics involved are typical game elements not unique to LoL.
We also believe that with the growing trend of gamification (e.g., as applied to citizen science) our findings have broader application than traditional entertainment gaming.
In fact, some gamified citizen science projects have already experience low level toxicity~\cite{foldit}.
As designers and scientists import more and more gaming elements into their systems, they will most assuredly be accompanied by less desirable elements of gaming culture, such as toxic behavior.

We also see similarities between LoL and online communities.
Online communities, e.g., Reddit, allow anonymous user identities.
They use unique user names which are not linked to real identity.
Although there are differences between forms of toxic playing and cyberbullying in online communities, the disconnect between real and virtual world is a common root that both cyberbullying and toxic playing stem from.

Beyond gaming, our work deepens understanding of team conflicts in general.
This has great potential because teams are an essential building block of modern organizations.
Also, in the current globally distributed workspace, collaboration through an electronic channel is pervasive.
This often results in a goal-oriented group that lacks \emph{social} connections.
Such settings can accelerate the tendency to blame others, further escalated by the individuals' unfamiliarity with remote partners~\cite{cramton2001mutual}.
Competitive games in general, and the LoL Tribunal in particular, are thus a valuable asset to capture group conflicts in the virtual space and test effective treatments for them.
We believe that solutions for toxic play in team competition online games could have huge impact for real-world scenarios, and not just virtual spaces.

For the overall CHI community, while ethnography is traditionally considered as one of the best methods to understand people, we show that studying human behavior with big data and testable hypotheses works well.
We hope that this helps accelerate the discovery of big data's value by the CHI community.

\subsection{Caveats and Limitations}

Although our findings confirm several theories on cyberbullying and toxic behavior, we note that there are limitations and caveats that must be considered.
First, although there is reason to believe that gamers generally behave the same in-game as they do in the real-world~\cite{boellstorff2008coming}, the fact remains that our dataset is drawn from a game.
At minimum, this means applying our results to other domains must be done carefully~\cite{keegan2011sic}.
Further, our findings are from LoL in particular, and there are domain specific concerns that might not be present in other games even within the same genre.
For example, SMITE is currently the 3rd most popular MOBA in the market, but completely lacks an all-talk chat mode.
Additionally, SMITE and Dota 2 (the 2nd most popular MOBA) both have integrated voice communication, which might intensify or soften cyberbullying with feelings of social presence.

Next, the social network structure among players, ``friend'' relationships, is not considered in this work due to lack of data.
We thus do not incorporate social relationship among specific players in building our hypotheses.
As recent studies reveal that playing together with friends influences performance and toxicity~\cite{Mason13,shores2014identification}, we believe that more detailed data including players' social networks would enable testing of more sophisticated hypotheses.

Finally, although our dataset is large-scale and quite rich, it is anonymized and thus prevents us from knowing how toxic players behaved before and after the matches were aggregated into their case and a decision was made.
We also lack knowledge of how other players in the game typically behave.
Thus, while we were able to provide significant support for several theories, there are questions for which answers continue to elude us.

\section{Acknowledgments}
This material is based on research sponsored by DARPA under agreement number FA8750-12-2-0107. The U.S. Government is
authorized to reproduce and distribute reprints for Governmental
purposes notwithstanding any copyright notation thereon.

\balance

% \section{References format}
% References must be the same font size as other body text.
% REFERENCES FORMAT
% References must be the same font size as other body text.

\bibliographystyle{acm-sigchi}
\bibliography{arxiv-main}
\end{document}